# The prescribed time sliding mode control for attitude tracking of spacecraft

Zhirun Chen[a], Xiaozhe Ju[b], Ziwei Wang[a], Qing Li[a,*]

[a]Department of Automation, Tsinghua University, Beijing 100084, China

[b]School of Astronautics, Harbin Institute of Technology, Harbin 150001, China

**Abstract**

With the development of the space missions, there are extensive missions in the demand of the prescribed time convergence. However, it's still a difficult work to combine the prescribed time method with the sliding mode control due to the infinite gain of the prescribed time method while approaching the prescribed time and two periods of sliding mode control. In this paper, a new prescribed time sliding mode control method is proposed for general systems with matched disturbances, from the second-order system to the high-order system. A novel sliding mode variable with explicit time term is designed for achieving the prescribed time convergence. More importantly, as time approaches the prescribed time, the singularity of control input can be avoided. Finally, this paper presents a disturbance observer based prescribed time sliding mode control method for attitude tracking of spacecraft and the efficiency of this method has been verified through the numerical simulations.

*Keywords*: prescribed time sliding mode control; spacecraft attitude control; singularity; matched disturbances.

## 1. Introduction

Recent years have witnessed the extensive researches of spacecraft attitude control. Several nonlinear control methods have been proposed for spacecraft attitude control, such as sliding mode control [1], robust $H_\infty$ control [24], backstepping control [4], inverse optimal control [5], geometric control [6], disturbance observer based control [7], and fault-tolerant control [2]-[3], etc. Although the above works tackle the tracking problem well, these methods hardly consider the convergence time of the spacecraft attitude control system. Extensive space missions, such as space rendezvous, space docking and space robots capture, should consider the convergence time. For example, in the space rendezvous mission, the target satellite should approach the desired attitude within a limited time window.

Nowadays massive researches have focused on achieving the finite-time convergence under the above demand, which can give a bound of the settling time. Terminal sliding mode control [8] is a typical finite-time control method, which gradually develops into nonsingular terminal sliding mode control [9] to avoid the singularity problem. Furthermore, nonsingular fast terminal sliding mode control [10], [25] is proposed to achieve faster convergence.

However, the settling time of finite-time control method depends on the initial states errors, such dependence can be relaxed with the fixed-time control [11]-[14]. However, the settling time of fixed-time method hasn't been in the form of an explicit function of control parameters, so it's still a tedious work to achieve the appointed convergence time by tuning the control parameters.

The aforementioned design difficulties can be relaxed by the prescribed time control [15]-[20]. Wang and Song [15]-[16] propose the prescribed time control method, which can achieve the settling time independent of initial states and control parameters, but its design process is complicated for high-order systems. Wang and Liang [17] extend the prescribed time control to the practically prescribed time control, but the singularity of the control input haven't been solved well. Pal [18] proposes a simplified prescribed time control scheme, but they hardly consider the singularity of control input as time approaches the prescribed time. Tran and Yucelen [19] study a new prescribed time control method for the perturbed dynamical systems based on a time transformation approach. However, in discrete form of this method, the sampling interval after transformation approach is very large while time approaches the prescribed time, which can't guarantee the stability of the practical system. Grag [20] proposes a prescribed time control method for a general class of affine nonlinear system with input constraints, but it also adopts a time transformation method with a stability problem while being applied in practical system.

This paper proposes a prescribed time sliding mode control for attitude tracking of spacecraft. The main contributions of this paper are stated as follows:

1) This paper proposes a prescribed time sliding mode control for general systems with matched disturbances, from the second-order system to the high-order system.
2) The proposed control method combines prescribed time with sliding mode control in a simpler form, and avoids the singularity while time approaches the prescribed time. Besides, some improvements are given while being applied in the practical systems.
3) The disturbance observer based prescribed time sliding mode control method is proposed for the attitude tracking of spacecraft with matched disturbances.

The remainder of this paper is arranged as follows. Section 2 introduces the formulation of spacecraft attitude tracking problem. In Section 3, the main contribution of this paper is stated. In Section 4, the disturbance observer based prescribed time sliding mode control is proposed for the spacecraft attitude control. Finally, conclusions are given in Section 5.

## 2. Problem statement and preliminaries

In this section, the spacecraft attitude dynamics model is described for problem statement. Besides, definitions of different convergence rates used in the following deduction process are specified for the ease of comprehension.

*2.1 Spacecraft attitude dynamics*

Assuming that spacecraft is a rigid body, the kinematic and dynamic equations of spacecraft attitude are described as follows [21]

$$\begin{cases} \dot{q}_1 = \dfrac{1}{2}T(q)w \\ \dot{q}_4 = -\dfrac{1}{2}q_1^T w \end{cases} \qquad (1)$$

and

$$J\dot{w} = -w^\times Jw + T + d \qquad (2)$$

where $q = [q_1^T, q]^T$ is the quaternion with $q_1 = [q_1, q_2, q_3]^T$, $w = [w_1, w_2, w_3]^T$ is the angular velocity vector, $J$ is the inertia matrix, $T = [T_1, T_2, T_3]$ is the control input torque, $d$ is the disturbance including solar radiation and magnetic effects, etc., and

$$T(q) = \begin{bmatrix} q_4 & -q_3 & q_2 \\ q_3 & q_4 & -q_1 \\ -q_2 & q_1 & q_4 \end{bmatrix} \qquad (3)$$

and the $q$ is constrained by $q_1^2 + q_2^2 + q_3^2 + q_4^2 = 1$.

Transforming the spacecraft attitude dynamic equations into the attitude tracking form:

$$\begin{cases} \dot{\varepsilon}_1 = 0.5T(q)w - \dot{q}_{1f} \\ \dot{\varepsilon}_4 = -0.5q_1^T w - \dot{q}_{4f} \\ J\dot{w} = -w^\times Jw + T + d \end{cases} \qquad (4)$$

where $\varepsilon_1 = q_1 - q_{1f}$ and $\varepsilon_4 = q_4 - q_{4f}$ represent the tracking errors.

In this paper, our ultimate aim is to develop a control law such that tracking errors can reach zeros within the prescribed time $t_f$.

*2.2 Preliminary*

Consider the system

$$\dot{x}(t) = f(x(t), t) \tag{5}$$

where $x(t) \in R^n$ is state, $f : R^n \times R_{\geq 0} \to R^n$ is a nonlinear function, and its initial condition is $x(0) = x_0$. Origin $x = 0$ is an equilibrium point of system (5).

**Definition 1.** (Finite-time stable, [12]): The origin of the system is called globally finite-time stable if it is globally asymptotically stable and any solution of system (5) converge to the origin at some finite time, $\forall t \geq t_0 + T(t_0, x_0)$, $x(t, x_0, t_0) = 0$, where $T$ is the settling time function.

**Definition 2.** (Fixed-time stable, [12]): The origin of the system is called fixed-time stable if it is globally finite-time stable and the settling time function is bounded, i.e., $\exists T_{max}$, s.t. $\forall x_0, t_0$, $T_{max} \geq T(t_0, x_0)$.

**Definition 3.** (Prescribed-time stable, [22]): The origin of the system is said to be prescribed-time stable if it is globally finite-time stable, and the settling time is a user-assignable finite constant, i.e., $\forall 0 < T_p \leq T_{max} < \infty$ ($T_p$ denotes the physically possible range), $T$ can be prescribed such that $T_p \leq T \leq T_{max}$.

**Assumption 1.** The disturbance $d$ of the spacecraft is bounded and satisfies $\|d\|_\infty \leq K$, where $K$ is a known positive constant. Besides, the derivative of the disturbance is bounded and satisfies $\|\dot{d}\|_\infty \leq c$, where $c$ is a known positive constant.

## 3. Main contribution

This section proposes the prescribed time sliding mode controllers for general systems with matched disturbances, from the second-order system to the high-order system.

*3.1 A second-order system case*

Consider the second-order system with matched disturbance

$$\begin{aligned} \dot{x}_1 &= x_2 \\ \dot{x}_2 &= f(x) + g(x)u + d \end{aligned} \tag{6}$$

where $x_1$, $x_2$ are states, $\mathbf{x} = [x_1, x_2]^T$, and $f$, $g$ are smooth function of $x_1$, $x_2$, and $u$ is the control input, $d$ is the matched disturbance. The disturbance $d$ satisfies the **Assumption 1**.

While $t < t_f$, a novel dynamic sliding variable is designed as

$$s = (t_f - t)x_2 + \eta x_1 \tag{7}$$

where $t_f$ is the prescribed time, and $\eta$ is a nonnegative constant.

The derivative of the sliding variable is:

$$\begin{aligned} \dot{s} &= (t_f - t)\dot{x}_2 - x_2 + \eta \dot{x}_1 \\ &= (t_f - t)\dot{x}_2 + (\eta - 1)x_2 \\ &= (t_f - t)(f(\mathbf{x}) + g(\mathbf{x})u + d) + (\eta - 1)x_2 \end{aligned} \tag{8}$$

While $t \geq t_f$, a classical sliding variable is selected as

$$s = x_1 + x_2 \tag{9}$$

In order to achieve prescribed time convergence, the control law is chosen as

$$u = \begin{cases} g(\mathbf{x})^{-1}\left(-\dfrac{\eta - 1}{t_f - t}x_2 - f(\mathbf{x}) - K\mathrm{sign}(s) - \dfrac{\eta s}{(t_f - t)^2}\right) & (t < t_f) \\ g(\mathbf{x})^{-1}\left(-f(\mathbf{x}) - K\mathrm{sign}(s) - K_1 s - x_2\right) & (t \geq t_f) \end{cases} \tag{10}$$

where $\eta > 2$, and $K_1$ is a positive constant.

**Theorem 1.** Consider the system (6) with matched disturbance, the control law (10) is adopted where $s = (t_f - t)x_2 + \eta x_1\ (t < t_f)$ and $s = x_1 + x_2\ (t \geq t_f)$, the system will be prescribed time stable.

**Proof.** The proof can be divided into two parts:

1) While $t < t_f$, design a Lyapunov function:

$$V = \frac{1}{2}s^2 \tag{11}$$

Its derivative is

$$\begin{aligned}\dot{V} &= s\dot{s} \\ &= s\left[(t_f - t)\left[f(\boldsymbol{x}) + g(\boldsymbol{x})\left[g(\boldsymbol{x})^{-1}\left(-\frac{\eta-1}{t_f-t}x_2 - f(\boldsymbol{x}) - K\mathrm{sign}(s) - \frac{\eta s}{(t_f-t)^2}\right)\right] + d\right] + (\eta-1)x_2\right] \\ &\leq -\eta s^2/(t_f - t) \\ &= -2\eta V/(t_f - t)\end{aligned} \quad (12)$$

If $V \neq 0$, then $\dot{V} < 0$, we transform the above inequality into the following form

$$\frac{\dot{V}}{V} \leq -\frac{2\eta}{(t_f - t)} \quad (13)$$

Then integrate the both sides of inequality (13)

$$\int_0^t \frac{\dot{V}}{V} dt \leq \int_0^t -\frac{2\eta}{(t_f - t)} dt \quad (14)$$

The following inequality will be obtained

$$V \leq C(t_f - t)^{2\eta} \quad (15)$$

where $C = V_0 / (t_f)^{2\eta}$. Consequently, $|s| \leq \sqrt{2C}(t_f - t)^{\eta}$.

Considering the sliding variable constraint, then

$$\left|(t_f - t)x_2 + \eta x_1\right| \leq \sqrt{2C}(t_f - t)^{\eta} \quad (16)$$

Transform Equation (16) for integration

$$\left|\frac{(t_f - t)^{\eta} \dot{x}_1 + \eta(t_f - t)^{\eta-1} x_1}{(t_f - t)^{2\eta}}\right| \leq \frac{\sqrt{2C}}{(t_f - t)} \quad (17)$$

From (17), we have

$$\left|d\left(\frac{x_1}{(t_f - t)^{\eta}}\right) \middle/ dt\right| \leq \frac{\sqrt{2C}}{(t_f - t)} \quad (18)$$

Let us integrate both sides of inequality (18)

$$\int_0^t -\frac{\sqrt{2C}}{(t_f-t)}dt \leq \int_0^t \left(d\left(\frac{x_1}{(t_f-t)^\eta}\right)\bigg/dt\right)dt \leq \int_0^t \frac{\sqrt{2C}}{(t_f-t)}dt \tag{19}$$

From (19), we have

$$\begin{aligned} C_1(t_f-t)^\eta + C_2(t_f-t)^\eta \ln(t_f-t) \leq x_1 \\ x_1 \leq C_3(t_f-t)^\eta + C_4(t_f-t)^\eta \ln(t_f-t). \end{aligned} \tag{20}$$

where $C_1, C_2, C_3, C_4$ are constants. From (16), we have

$$|x_2| \leq \sqrt{2C}(t_f-t)^{\eta-1} + \eta|x_1|/(t_f-t) \tag{21}$$

From (20), we have

$$|x_2| \leq C_5(t_f-t)^{\eta-1} + C_6(t_f-t)^{\eta-1}\ln(t_f-t) \tag{22}$$

where $C_5, C_6$ are constants.

If $V=0$, then $\dot{V} \leq 0$, so $V$ will remain at zero, and $s=0$. From (16) to (22), we have the similar inequality as (22).

Considering the control input with (16) and (22), If $\eta > 2$, then $\lim_{t \to t_f^-}(t_f-t)^{\eta-2}\ln(t_f-t)=0$, so the control input will be bounded, which avoids the singularity though the gain will approach infinity. What's more, $x_1$ and $x_2$ will approach zeros within the prescribed time.

2) While $t \geq t_f$, design a Lyapunov function $V = 0.5s^2$, then its derivative is

$$\dot{V} \leq -K_1 s^2 = -2K_1 V \tag{23}$$

From part 1, while $t=t_f$, the system reaches the sliding mode surface $s = x_1 + x_2$ and $x_1 = x_2 = 0$. According to the Lasalle's Invariance Principle, the system will keep in the origin while $t \geq t_f$.

In conclusion, the system will be prescribed time stable.

**Remark 1.** Considering the practical implication of this control method, the infinity gain is the main problem under a finite sampling frequency. Consequently we improve the above control law by the following approach

$$u = \begin{cases} g(x)^{-1}\left(-\dfrac{\eta-1}{t_f - t}x_2 - f(x) - Ksign(s) - \dfrac{\eta s}{(t_f - t)^2}\right) & \left(t < t_f - \delta\right) \\ g(x)^{-1}\left(-f(x) - Ksign(s) - K_1 s - x_2\right) & \left(t \geq t_f - \delta\right) \end{cases} \quad (24)$$

where $\delta > 0$ is a small positive constant, for example, we can choose $\delta = 0.1$ or $\delta = 0.01$ in different situations. In order to guarantee the tracking accuracy in the prescribed time, we can just increase the $\eta$, which will increase the convergence rate according to the Equation (16) and (22). On the other side, we should consider the control input constraint of the practical system, we can just decrease the $\eta$ to reduce the initial control input demand, so there will be a trade off between the tracking accuracy and control input constraint.

**Remark 2.** Considering the chattering problem of the high-frequency switching, the different disturbance observer, such as nonlinear disturbance observer, finite-time disturbance observer and prescribed-time disturbance observer, can be used to attenuate chattering.

The simulation of this second order system is carried out with $f = 0$, $g = 1$, $d = 0.01\sin(t)$, $x_1(0) = 5$, $x_2(0) = 3$, $\eta = 3$ and $t_f = 5$ seconds. The state history and control input are shown in the Fig. 1. It can be verified that origin is stabilized within a prescribed time $t_f$.

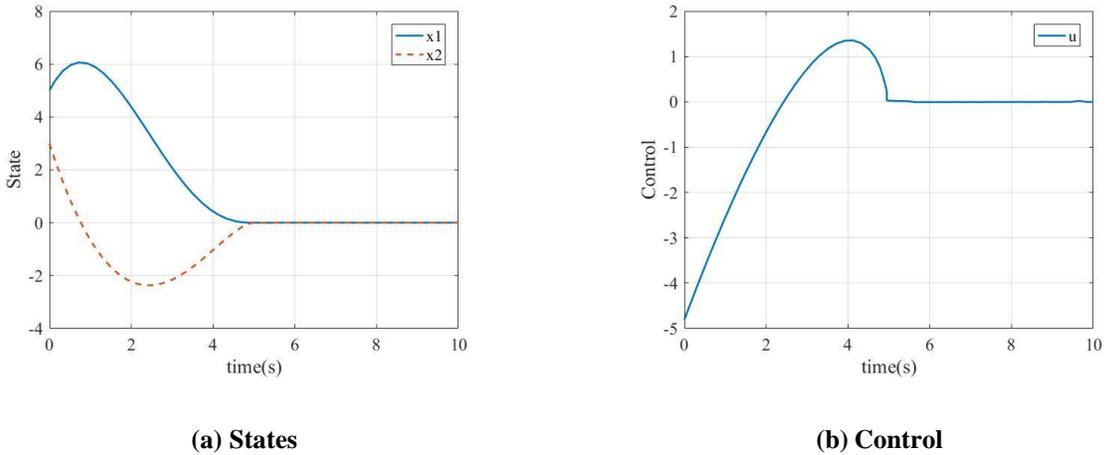

(a) States          (b) Control

**Fig. 1. States and control for second order system**

*3.2 A practical second-order system case*

Considering the practical second-order system with matched disturbances

$$\begin{aligned} \dot{x}_1 &= F(x) \\ \dot{x}_2 &= H(x) + G(x)u + d \end{aligned} \quad (25)$$

where $x_1 \in R^n$, $x_2 \in R^n$ are states, $x = [x_1^T, x_2^T]^T$, and $F$, $G$, $H$ are smooth function of $x_1$, $x_2$, and $u \in R^n$ is the control input, $d \in R^n$ is the matched disturbance. The disturbance $d$ satisfies the **Assumption 1**.

While $t < t_f$, a novel dynamic sliding variable is designed as

$$s = (t_f - t)\dot{x}_1 + \eta x_1 \tag{26}$$

where $t_f$ is the prescribed time, and $\eta$ is a nonnegative constant.

The derivative of the sliding variable is:

$$\begin{aligned}\dot{s} &= (t_f - t)\left(\frac{\partial F}{\partial x_1}\dot{x}_1 + \frac{\partial F}{\partial x_2}\dot{x}_2\right) - \dot{x}_1 + \eta \dot{x}_1 \\ &= (t_f - t)\left(\frac{\partial F}{\partial x_1}F(x) + \frac{\partial F}{\partial x_2}(H(x) + G(x)u + d)\right) + (\eta - 1)\dot{x}_1\end{aligned} \tag{27}$$

While $t \geq t_f$, a classical sliding variable is selected as

$$s = x_1 + \dot{x}_1 \tag{28}$$

The similar control law can be designed as

$$u = \begin{cases} -\left(\dfrac{\partial F}{\partial x_2}G(x)\right)^{-1}\left(\dfrac{\eta-1}{t_f-t}F(x) + \dfrac{\partial F}{\partial x_2}H(x) + K\left\|\dfrac{\partial F}{\partial x_2}\right\|sign(s) + \dfrac{\partial F}{\partial x_1}F(x) + \dfrac{\eta s}{(t_f-t)^2}\right) & (t < t_f) \\ -\left(\dfrac{\partial F}{\partial x_2}G(x)\right)^{-1}\left(\dfrac{\partial F}{\partial x_2}H(x) + K\left\|\dfrac{\partial F}{\partial x_2}\right\|sign(s) + K_1 s + \dfrac{\partial F}{\partial x_1}F(x)\right) & (t \geq t_f) \end{cases} \tag{29}$$

where $\eta > 2$, and $K = diag(K, K, ..., K) \in R^{n \times n}$, $K_1$ is a positive matrix.

**Theorem 2.** Consider the system (25) with matched disturbance, the control law (29) is adopted where $s = (t_f - t)\dot{x}_1 + \eta x_1 \ (t < t_f)$ and $s = x_1 + \dot{x}_1 \ (t \geq t_f)$, the system will be prescribed time stable.

**Proof.** The proof can be divided into two parts:

1) While $t < t_f$, design a Lyapunov function:

$$V = \frac{1}{2}s^T s \tag{30}$$

Its derivative is

$$\dot{V} = s^T \dot{s}$$
$$\leq -\frac{\eta s^T s}{(t_f - t)} \tag{31}$$
$$= -\frac{2\eta V}{(t_f - t)}$$

From (13) and (15), we have $\|s\|_2 \leq \sqrt{2C}(t_f - t)^\eta$, as $\|s\|_\infty \leq \|s\|_2$, then we have

$$\left|(t_f - t)\dot{x}_{1i} + \eta x_{1i}\right| \leq \|s\|_\infty \leq \|s\|_2 \leq \sqrt{2C}(t_f - t)^\eta \tag{32}$$

where $x_{1i}$ is the i-th component of $x_1$ and $i = 1, 2, 3$.

From (16) to (22), in the similar way we have

$$|\dot{x}_{1i}| \leq C_{5i}(t_f - t)^{\eta-1} + C_{6i}(t_f - t)^{\eta-1} \ln(t_f - t) \tag{33}$$

where $C_{5i}, C_{6i}$ are constants.

Considering the control input with (32) and (33), the control input will be bounded, which avoids the singularity though the gain will approach infinity. What's more, $x_1$ and $x_2$ will approach zeros within the prescribed time.

2) While $t \geq t_f$, design a Lyapunov function $V = 0.5 s^T s$, then we have

$$\dot{V} \leq -s^T K_1 s = -2\lambda_{\min}(K_1) V \tag{34}$$

where $\lambda_{\min}(K_1)$ is the smallest eigenvalue of $K_1$. In the same way of Section 3.1, the system will keep in the origin while $t \geq t_f$.

In conclusion, the system will be prescribed time stable. Considering the practical application of this control method and chattering problem, we can improve them in the same way as in the **Remark 1** and **Remark 2**.

The simulation of this part adopts the spacecraft dynamic model, which will be shown in the Section. 4.

*3.3 A general high-order system case*

Consider a general high-order system with matched disturbance

$$\dot{x}_i = x_{i+1} \quad (i=1,2,...,n-1)$$
$$\dot{x}_n = f(\boldsymbol{x}) + g(\boldsymbol{x})u + d \tag{35}$$

where $x_i\,(i=1,2,n)$ is state, $\boldsymbol{x} = [x_1, x_2,...x_n]^T$, and $f$, $g$ are smooth function of $\boldsymbol{x}$, and $u$ is the control input, $d$ is the matched disturbance. The disturbance $d$ satisfies the **Assumption 1**.

While $t < t_f$, a novel sliding variable is designed as

$$s = \frac{d^{n-1}\left(\dfrac{x_1}{(t_f-t)^{\eta}}\right)}{dt^{n-1}}(t_f-t)^{\eta+n-1} \tag{36}$$

where $t_f$ is the prescribed time, and $\eta$ is a nonnegative constant. We expand (35) as follows

$$s = \sum_{i=0}^{n-1} c_i (t_f-t)^i x_1^{(i)} \tag{37}$$

where $c_i\,(i=0,1,...,n-1)$ is the expansion parameter, $x_1^{(i)}$ is the i-th derivative of the $x_1$, and the derivative of the sliding variable is

$$\dot{s} = \sum_{i=0}^{n-1}\left(c_i - c_{i+1}(1+i)\right)(t_f-t)^i x_1^{(i+1)} + c_{n-1}(t_f-t)^{n-1} x_1^{(n)} \tag{38}$$

While $t \geq t_f$, let us choose a classical sliding variable

$$s = x_n + \sum_{i=1}^{n-1} a_i x_i \tag{39}$$

where $a_i\,(i=1,2,...,n)$ are constants, and they are assigned such that matrix

$$A = \begin{bmatrix} 0 & 1 & 0 & 0 \\ 0 & 0 & ... & ... \\ ... & ... & 0 & 1 \\ -a_1 & -a_2 & ... & -a_{n-1} \end{bmatrix}$$

is Hurwitz.

In order to achieve prescribed time convergence, the control law is designed as follows

$$u = \begin{cases} g(\boldsymbol{x})^{-1}\left(-\dfrac{\sum_{i=0}^{n-1}(c_i - c_{i+1}(1+i))(t_f - t)^i x_{i+2}}{c_{n-1}(t_f - t)^{n-1}} - f(\boldsymbol{x}) - K\mathrm{sign}(s) - \dfrac{(\eta + n - 2)s}{(t_f - t)^n}\right) & (t < t_f) \\ g(\boldsymbol{x})^{-1}\left(-f(\boldsymbol{x}) - K\mathrm{sign}(s) - K_1 s - \sum_{i=1}^{n-1} a_i x_{i+1}\right) & (t \geq t_f) \end{cases} \quad (40)$$

where $\eta > n$, $K_1$ is a positive constant.

**Theorem 3.** Consider the system (35) with matched disturbance, the control law (40) is adopted where $s = d^{n-1}\left(x_1/(t_f - t)^\eta\right)/dt^{n-1}$ $(t < t_f)$ and $s = x_n + \sum_{i=1}^{n-1} a_i x_i$ $t \geq t_f$, the system will be prescribed time stable.

**Proof.** The proof can be divided into two parts:

1) While $t < t_f$, design a Lyapunov function

$$V = \frac{1}{2}s^2 \tag{41}$$

Its derivative is

$$\begin{aligned} \dot{V} &= s\dot{s} \\ &\leq -\frac{2(\eta + n - 2)V}{(t_f - t)} \end{aligned} \tag{42}$$

From (13) and (15), we have $|s| \leq \sqrt{2C}(t_f - t)^{\eta + n - 2}$, then we have

$$\left|\left(d^{n-1}\left(\frac{x_1}{(t_f - t)^\eta}\right)\bigg/dt^{n-1}\right)(t_f - t)^{\eta + n - 1}\right| \leq \sqrt{2C}(t_f - t)^{\eta + n - 2} \tag{43}$$

Transform (43) into the following form

$$\left|d^{n-1}\left(\frac{x_1}{(t_f - t)^\eta}\right)\bigg/dt^{n-1}\right| \leq \frac{\sqrt{2C}}{(t_f - t)} \tag{44}$$

Before integrating both sides of inequality (44), two ordinary integral formula are shown as follows

$$\int_0^t \ln(t_f - t)dt = t\ln(t_f - t)\Big|_0^t + \int_0^t \frac{t}{(t_f - t)}dt$$
$$= t\ln(t_f - t) - t + t_f \ln(t_f) - t_f \ln(t_f - t) \tag{45}$$

$$\int_0^t t^{n-1} \ln(t_f - t)dt = \frac{1}{n}\left(t^n \ln(t_f - t)\Big|_0^t + \int_0^t \frac{t^n}{(t_f - t)}dt\right)$$
$$= \frac{1}{n}\left(t^n \ln(t_f - t) + \sum_{i=0}^n \frac{a_i(t_f)^n - a_i(t_f - t)^n}{n} + a_0(\ln(t_f) - \ln(t_f - t))\right) \tag{46}$$

where $a_i (i = 0, 1, ..., n)$ are constants.

From (45) and (46), it's easy to obtain the following equality by integrating n-1 times on both sides of inequality (44)

$$|x_1| \leq C_1(t_f - t)^{\eta} + C_2(t_f - t)^{\eta} \ln(t_f - t) \tag{47}$$

where $C_1, C_2$ are constants. Consequently let us integrate n-2 times on both sides of inequality (44), we have

$$|(t_f - t)x_2 + \eta x_1| \leq (t_f - t)^{\eta+1}(C_3 \ln(t_f - t) + C_4) \tag{48}$$

It follows that

$$|x_2| \leq (t_f - t)^{\eta}(C_3 \ln(t_f - t) + C_4) + (t_f - t)^{\eta-1}(C_5 \ln(t_f - t) + C_6) \tag{49}$$

where $C_3, C_4, C_5, C_6$ are constants.

Similarly, we integrate both sides of inequality (44) n-i times, the following equality can be obtained

$$|x_i| \leq \sum_{k=0}^{i}(t_f - t)^{\eta-k+1}(D_{2k} \ln(t_f - t) + D_{2k+1}) \tag{50}$$

where $D_{2k}, D_{2k+1}$ are constants.

Considering the control input with (43) and (50), If $\eta > n$, then $\lim_{t \to t_f^-}(t_f - t)^{\eta-n} \ln(t_f - t) = 0$, so the control input will be bounded, which avoids the singularity though the gain will approach infinity. What's more, $x_i (i = 1, ..., n)$ will approach zeros within the prescribed time.

2) While $t \geq t_f$, design a Lyapunov function $V = 0.5s^2$, then we have

$$\begin{aligned}\dot{V} &\leq s\dot{s} \\ &\leq -Kss \\ &\leq -2KV\end{aligned} \quad (51)$$

In the same way of Section 3.1, according to the Lasalle's Invariance Principle, the system will keep in the origin while $t \geq t_f$.

In conclusion, the system will be prescribed time stable.

**Remark 3.** Considering the practical application of this control method and chattering problem, we can improve them in the same way as in the **Remark 1** and **Remark 2**. The improved control method is shown as follows

$$u = \begin{cases} g(\boldsymbol{x})^{-1}\left(-\dfrac{\sum_{i=0}^{n-1}(c_i - c_{i+1}(1+i))(t_f - t)^i x_{i+2}}{c_{n-1}(t_f - t)^{n-1}} - f(\boldsymbol{x}) - K\text{sign}(s) - \dfrac{(\eta + n - 2)s}{(t_f - t)^n}\right) & (t < t_f - \delta) \\ g(\boldsymbol{x})^{-1}\left(-f(\boldsymbol{x}) - K\text{sign}(s) - K_1 s - \sum_{i=1}^{n-1} a_i x_{i+1}\right) & (t \geq t_f - \delta) \end{cases} \quad (52)$$

The simulation of the third order system is carried out with $f = 0$, $g = 1$, $d = 0.01\sin(t)$, $x_1(0) = 5$, $x_2(0) = 3$, $x_3(0) = 2$, $\eta = 4$ and $t_f = 5$ seconds. The state history and control input are shown in the Fig. 2. It can be verified that origin is stabilized within a prescribed time $t_f$.

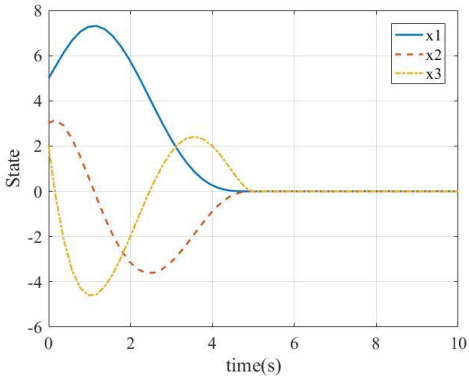
(a) States

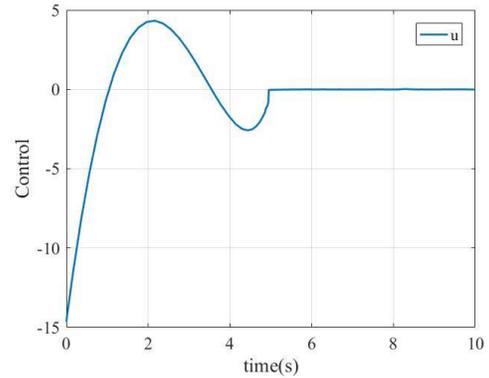
(b) Control

**Fig. 2. States and control for third order system**

## 4. Applications to the spacecraft attitude control

This section proposes the nonlinear disturbance observer based prescribed time sliding mode controller for spacecraft attitude control. Besides, two simulation cases are presented to verify the efficiency of the proposed method.

*4.1 The disturbance observer based prescribed time controller*

The disturbance $d$ of the spacecraft satisfies the **Assumption 1**, then disturbance observer is designed as

$$\begin{cases} \dot{z} = -l(w)\left[-J^{-1}w^{\times}Jw + J^{-1}T + J^{-1}\hat{d}\right] \\ \hat{d} = z + \lambda(w) \end{cases} \quad (53)$$

where $l(w) = \partial \lambda(w)/\partial w$ and $l(w)$ is a positive diagonal matrix.

The disturbance observer estimation error is defined as $e_d = \hat{d} - d$, then we have

$$\dot{e}_d = -l(w)J^{-1}e_d - \dot{d} \quad (54)$$

The disturbance estimation error is bounded by

$$\begin{aligned} \|e_d(t)\| &= \left\| e^{-l(w)J^{-1}t} e_d(0) - \int_0^t (e^{-l(w)J^{-1}(t-\tau)} \dot{d}) d\tau \right\| \\ &\leq e^{-l_m t}\|e_d(0)\| + c\int_0^t e^{-l_m(t-\tau)} dt \\ &\leq \frac{c}{l_m} + e^{-l_m t} c_1 \end{aligned} \quad (55)$$

where $c_1 = \|e_d(0)\| - c/l_m$ and $l_m$ is the minimum value of the matrix $l(w)J^{-1}$ diagonal elements. Let us define a positive variable $K_2$ as $K_2 = c/l_m + e^{-l_m t} c_1$.

The prescribed time sliding mode control method is designed according to the Section 3.2. While $t < t_f - \delta$, the novel sliding variable is selected as

$$s = (t_f - t)\dot{\varepsilon}_1 + \eta \varepsilon_1 \quad (56)$$

The control law is chosen as

$$U = w^{\times}Jw - \hat{d} - (T(q)J^{-1})^{-1}\|T(q)J^{-1}\|K_2 sign(s) - \frac{2(T(q)J^{-1})^{-1}(\eta-1)(\dot{q}_1 - \dot{q}_{1f})}{t_f - t}$$
$$-\frac{2(T(q)J^{-1})^{-1}\eta s}{(t_f - t)^2} - (T(q)J^{-1})^{-1}\left(\dot{T}(q)w - 2\ddot{q}_{1f}\right) \quad (57)$$

where $\eta > 2$, and

$$\dot{T}(q) = \begin{bmatrix} \dot{q}_4 & -\dot{q}_3 & \dot{q}_2 \\ \dot{q}_3 & \dot{q}_4 & -\dot{q}_1 \\ -\dot{q}_2 & \dot{q}_1 & \dot{q}_4 \end{bmatrix} \quad (58)$$

While $t \geq t_f - \delta$, the classical sliding variable is selected as

$$s = \dot{\varepsilon}_1 + \varepsilon_1 \quad (59)$$

The control law is selected as

$$U = w^{\times}Jw - \hat{d} - (T(q)J^{-1})^{-1}\|T(q)J^{-1}\|K_2 sign(s) - (T(q)J^{-1})^{-1}\left(\dot{T}(q)w - 2\ddot{q}_{1f} + 2\dot{\varepsilon}_1 + K_1 s\right) \quad (60)$$

where $K_1$ is a positive constant.

*4.2 Simulation results*

In order to illustrate the effectiveness of the proposed prescribed time sliding mode controllers for the spacecraft attitude tracking, two simulation cases will be presented in this section.

The spacecraft parameters are chosen as follows [23]:

$$J = diag(10,12,14)kg \cdot m^2, \ q(0) = \left[\sqrt{2}/3, -1/3, \sqrt{3}/3, \sqrt{3}/3\right]^T,$$

$$q(t_f) = [0,0,0,1]^T, \ d = [0.01\sin(0.1t), 0.01\sin(0.1t), 0.01\sin(0.1t)]^T,$$

$$K_1 = 1, \ l(w) = diag(10,10,10).$$

**A. Case I**

This case adopts the different prescribed time $t_f = 30s$ and $t_f = 40s$ for comparison, other parameters are chosen as $\delta = 0.05$ and $\eta = 3$. The attitude tracking history and control torque of this case are shown from Fig. 3 to Fig. 6.

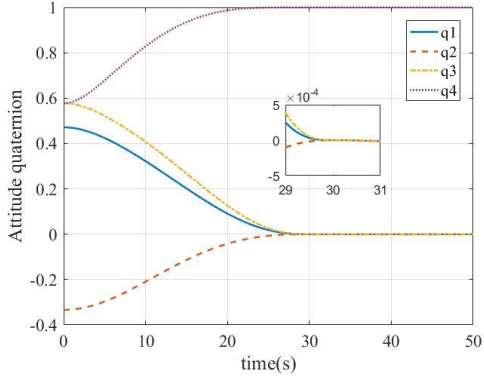
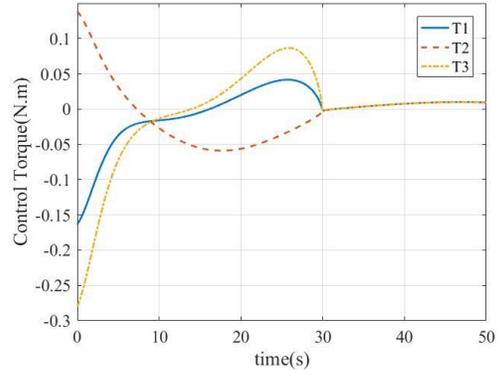

**Fig. 3.** $t_f = 30s$ **Attitude Tracking**    **Fig. 4.** $t_f = 30s$ **Control Torque**

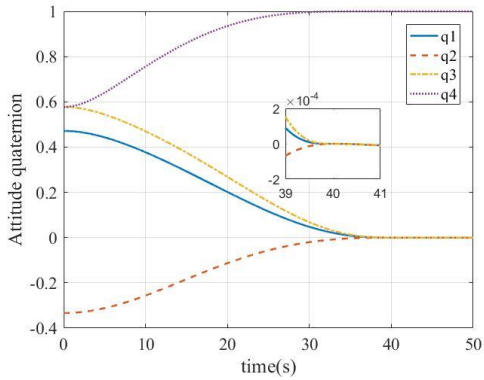
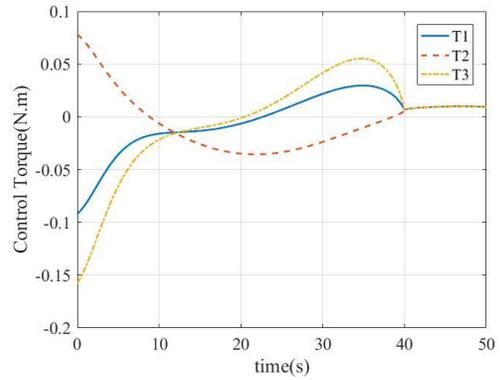

**Fig. 5.** $t_f = 40s$ **Attitude Tracking**    **Fig. 6.** $t_f = 40s$ **Control Torque**

It is observed from the Fig.3 to Fig.6 that attitude quaternion can approach the desired output with a 10E-4 magnitude tracking errors at the prescribed time. Furthermore, if the prescribed time is shorter, the initial control torque is larger.

**B. Case II**

This case adopts different control parameters $\eta = 3$ and $\eta = 5$ for comparison, other parameters are chosen as $t_f = 35s$ and $\delta = 0.05$. The attitude tracking history and control torque of this case are shown from Fig. 7 to Fig. 10.

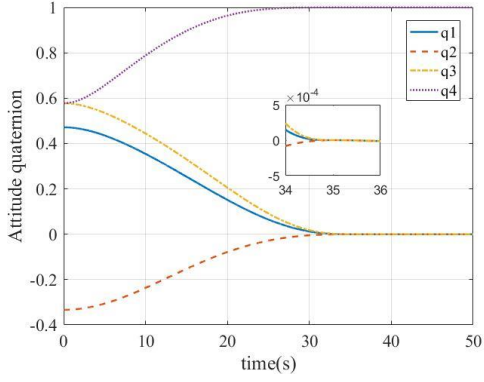
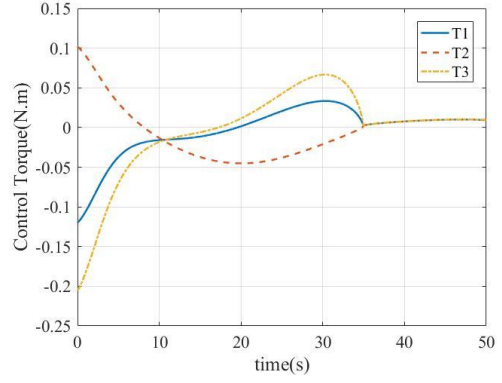

**Fig. 7.** $\eta = 3$ **Attitude Tracking**     **Fig. 8.** $\eta = 3$ **Control Torque**

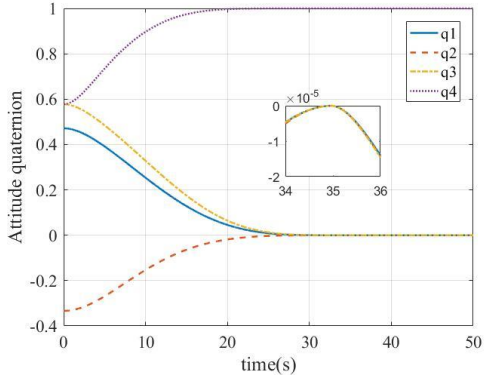
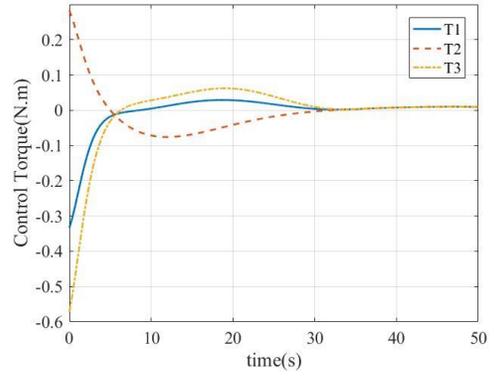

**Fig. 9.** $\eta = 5$ **Attitude Tracking**     **Fig. 10.** $\eta = 5$ **Control Torque**

It is observed from the Fig.7 to Fig.10 that attitude quaternion can approach the desired output at the prescribed time with a 10E-4 magnitude tracking errors while $\eta = 3$ and a 10E-5 magnitude tracking errors while $\eta = 5$. Furthermore, if the parameter $\eta$ is larger, the initial control torque is larger and the output converges faster.

## 5. Conclusion

This paper proposes a disturbance observer based prescribed time sliding mode control method for spacecraft attitude tracking with matched disturbance. The proposed control method combines prescribed time with sliding mode control in a simpler form, and avoids the singularity. Besides, this control method is generalized for the high-order systems. In the future, we should consider the control input constraints and optimality while being applied in the practical system.

**Acknowledgments**

This work is sponsored by the National Natural Science Foundation of China No. 61771281, the "New generation artificial intelligence" major project of China No. 2018AAA0101605, the 2018 Industrial Internet innovation and development project, and Tsinghua University initiative Scientific Research Program.

**Conflict of interests**

The authors declare that there is no conflict of interest regarding the publication of this paper.